# Different approaches to unveil biomolecule configurations and their mutual interactions

R. Benedetti[a], F. Bajardi[b,c], S. Capozziello[b,c,d], V. Carafa[a], M. Conte[a], M. R. Del Sorbo[e], A. Nebbioso[a], M. Singh[b], H. G. Stunnenberg[f], M. Valadan[b,c], L. Altucci[a], and C. Altucci[b,c]∗

[a]*Dipartimento di Biochimica, Biofisica e Patologia Generale, Università degli Studi della Campania "L. Vanvitelli", Napoli, Italy;* [b]*Dipartimento di Fisica "Ettore Pancini", Università degli Studi di Napoli "Federico II", Napoli, Italy;* [c]*Istituto Nazionale di Fisica Nucleare (INFN), Sez. Di Napoli, Napoli, Italy;* [d]*Gran Sasso Science Institute, L'Aquila, Italy;* [e]*Istituto Statale d'Istruzione Superiore "Leonardo da Vinci", Poggiomarino (NA), Italy;* [f]*Department of Molecular Biology, NCMLS, Radboud University, Nijmegen, the Netherlands*

*corresponding author: carlo.altucci@unina.it, tel: 081-679286

# Different approaches to unveil biomolecule configurations and their mutual interactions

A novel technique was demonstrated that overcome important drawbacks to crosslink cells by irradiation with ultrashort UV laser pulses (L-crosslinking). To use this technique coupled to Chromatin ImmunoPrecipitation (ChIP) in a high throughput context, a pre-screening fast method needs to be implemented to set up suitable irradiation conditions of the cell sample for efficient L-crosslinking with no final and long ChIP analysis. Here a fast method is reported where living human cells have been first transfected with a vector coding for Estrogen Receptor α (ERα), linked to Green Florescent protein (ERα-GFP), so that the well-known interaction between the Estrogen Receptor Elements (ERE) region of the cell DNA and the ERα protein can be detected by studying the fluorometric response of the irradiated cells. The damage induced to cells by UV irradiation is characterized by looking at DNA integrity, proteins stability and cellular viability. A second novel approach is presented to analyze or re-visit DNA and RNA sequences and their molecular configurations. This approach is based on methods derived from Chern-Simons super-gravity adapted to describe mutations in DNA/RNA strings, as well as interactions between nucleic acids. As a preliminary case we analyze the KRAS human gene sequence and some of its mutations. Interestingly, our model shows how the Chern-Simons current are capable to characterize the mutations within a sequence, in particular giving a quantitative indication of the mutation likelihood.

Keywords: chromatin analysis; molecular biophysics; optical physics, Chern-Simons currents, genetic code.

**Introduction**

Studying and understanding the interactions between complex biomolecules in their native context and their configurations, possibly related to new ongoing interactions, is one of the hottest topics in the present epoch, involved in very many disciplines of science from genetics, proteomics, molecular biology, to chemistry and physics. Just to mention two noticeable cases where interactions between biomolecules play a major role we first cite here the DNA-protein interactions which is a pivotal event governing cellular functions, such as transcriptional regulation, chromosome maintenance, replication and DNA repair (Polo and Jackson 2011) and it is critical in development and environmental adaptation. Another crucial scenario is represented by the attempt to explain the genetic code of living organisms (Stanley et al. 1996; Gamow 1954) and viruses (Rossmann and Johnson 1989), a formidable task with immense potential fallout, that implies, to some extent, the capability of predicting gene mutations and to correctly represent the genetic alphabet. Here, for instance, interactions of nucleic acids with proteins are at fundament of very basic mechanisms such as DNA-RNA transcription.

In such a fertile scientific context, in tumultuous development, methods imported from various different branches of physics have been used to model and/or experimentally test several aspects of the interactions between complex biomolecules in a multi-disciplinary approach.

Here we shall focus on two examples, pretty much different from each other, where physics come into play to provide, in the former case, an experimental tool to genetics and epigenetics to fix interactions between DNA and proteins in living cells, by instantaneously welding proteins bounded to DNA thanks to the laser-induced crosslinking (L-crosslinking). This technique is based on the well-known ability of UV radiation to stably crosslink DNA and proteins due to the efficient absorption of nucleobases in the 250-300 nm UV band (Alexander and Moroson 1962). In particular, UV laser pulses are capable to freeze very rapidly DNA-protein interactions, potentially acting as an instantaneous welder through irreversible photoreactions, thus allowing, at least in principle, real-time investigation of the temporal and spatial binding of proteins on DNA. This technique was first proved effective in *in-vitro* experiments (Russmann et al. 1997; Russmann et al. 1998) and more recently in experiments on living human cells, both normal and cancer (Altucci et al. 2012). Lately, L-crosslinking has been applied in large scale analysis of the interaction between DNA and proteins in human cancer cells, proving to act as a so-called zero-length crosslinker, i.e. it is excellent in detecting direct interactions, that involve molecules in touch by their molecular orbital, and nearly blind to detect instead indirect interactions (Nebbioso et al. 2017). This characteristic, combined with the use of ordinary crosslinking which is equally sensitive to detect both direct and indirect interactions, offers the exciting and unprecedented tool for studying short-lived and dynamic associations in gene regulatory networks and epigenetic mechanisms (Nagaich et al. 2004). Therefore, L-crosslinking is particularly indicated for high throughput large scale analysis of the interactions involving the entire genome; this type of massive analysis on an (epi)genome-wide scale is based on the use of sophisticated and expensive techniques such as chromatin immunoprecipitation (ChIP) coupled with sequencing (ChIP-seq) (Adams et al. 2012; Martens et al. 2011; Kaspi et al. 2012). Besides being sophisticated and expensive, ChIP-seq can also take long, thus demanding for a pre-screening method capable to quickly discriminate the most interesting conditions for DNA-protein interactions to be identified and studied in living cells, amongst many possible different cases. Here we focus, in the first part of the paper, on description and test of a possible fast screening method, based on the idea to tag with a fluorescent label the interaction to be probed and detected. By this way the pre-screening phase of the entire experiment, that allows the experimentalist to focus on the only interesting conditions discarding the remaining, just ends up with a fluorometric analysis, simple, fast, cheap and pretty reliable.

In the second part of the study, inspired by the strongly multidisciplinary view that the analysis of biomolecule interactions requires, we turn our attention to a completely different approach to describe the configuration of complex biomolecules, possibly in their native context. The analysis, which transforms series of amino acids in pure numbers, can be used to gain important information, such as locating the most likely specific locus within the large biomolecule for interactions between DNA and RNA and/or proteins, or to predict the position of the mutations within a given DNA/RNA string. This highly innovative method mainly deals with DNA-DNA or DNA-RNA interactions, such as docking between these large and fundamental biomolecules, and can be used also to predict mutations position into DNA strings based on likelihood quantitative criteria.

This method uses the formalism of topological field theories, in particular of the so-called Chern-Simons Gravity (Zanelli 2008). This new theory of the gravity is one of the most powerful alternatives to General Relativity. General Relativity, in fact, has several troubles, there including a hard match with a complete and coherent theory of Quantum Gravity (DeWitt 1967; Deser et al. 1982) and the impossibility to predict the today observed universe exponential expansion as well as the early-time universe inflation (Carrol et al. 2004; Kolb and Turner 1990). With the aim to solve the above mentioned and other shortcomings, several alternative theories of gravity have been constructed; among these, one of the most promising is the Chern-Simons theory which, treating gravity under the same formalism as other fundamental interactions, can successfully overcome at least the high-energy issues of Einstein's theory and contains General Relativity as a particular limit.

Chern-Simons formalism has been recently adopted with success to address some fundamental and stimulating problems in biology such as the presence of knotted DNA and their interactions with proteins (Dobrowski-Tumanski and Sulkowska 2017), and the unknotted RNA folding with the role of knotted protein in codon correction of RNA in methyl transfer (Lachner and Jenuwein 2002). Knots, that occur so often in interacting structures of the most recent biology, are four dimensional topological objects, embedded in three dimensions, very much suited to be treated within Chern-Simons formalism. They were used for long time in loop-quantum gravity and superconductor theory. For instance, a theoretical route has been traced for the general equation to solve knotted protein folding (Milo and Zewail 2012) by using the Wilson loop operator in loop-quantum gravity for gene expression with a boundary phase condition. A Chern-Simons-based theory should also support the problem of adaptive changing of docking curvature of knotted protein folding.

Here we use the formalism of Chern-Simons gravity in three dimensions, since we describe sequences of triplets in nucleic acids; the considered space describes the configurations containing all the possible combinations among nitrogen bases; since we know that nitrogen bases combine with each other in triplets, such space will be three-dimensional. This has been recently realized in (Capozziello et al. 2018; Capozziello and Pincak 2018). Therefore, while the standard Chern-Simons gravity acts in the n-dimensional super-manifold constituted by space-time coordinates, in our application we only focus on the equivalent space of configuration of nitrogen bases. Here we focus on the analysis, by means of our Chern-Simons-based method, of genetic sequences belonging to the KRAS human gene, a gene that acts as an on/off switch in cell signaling. When KRAS functions normally, it controls cell proliferation. When it is mutated, signaling is disrupted. Thus, cells can continuously proliferate, and often develop into cancer (Kranenburg 2005; Hartman et al. 2012; Chiosea et al. 2011). We analyzed a particular region of KRAS human gene, introducing some known mutations leading to given disease; the computation of the Chern-Simons current relative to the original sequence and the mutated one, shows a discrepancy between the two strings.

Once introducing the mutations in a certain point, the point-like Chern-Simons current increases (or decreases) its value and this permits to compare the original and the mutated sequence. Being related to the curvature of the amino acid string, a significant difference of the Chern-Simons current value implies a difference of curvature between the two related configurations. In this case we considered only mutations by replacement: one single nitrogen basis is substituted with another one and the difference in curvature can be used to set up a likelihood criterion for certain mutation to take place. Gradients in the Chern-Simons current are associated to strongly peaked areas in terms of curvature that need a consistent amount of energy to be formed. Thus, we may infer that these regions will tend to evolve to a lower energy configuration due to a principle of minimum energy. Therefore, they might be the best candidates for having a mutation.

Furthermore, the method potentially allows addressing the intriguing and vast problem of docking between biomolecules while they interact. Within this context, the method suggests a natural criterion to evaluate the docking chance and configuration: in fact, we might in principle estimate the docking point, by comparing the areas with similar curvature, what is still under investigation.

The paper is organized as follows: the first section is dedicated to illustrate the fast screening protocol that we settled in view of high throughput L-crosslinking experiments, with its sub-section of materials and methods. A second section is devoted to the above-mentioned application of Chern-Simons-based method to the KRAS gene analysis.

**Fast pre-screening method for high throughput experiments based on L-crosslinking in living cells**

Crosslinking induction between DNA and proteins in living cells with fs-UV lasers, the so-called L-crosslinking, has been presented in the literature (Altucci et al. 2012; Nebbioso et al. 2017) as a useful tool primarily to study transient interactions in living cells. This methodology paves the way to substitute chemically inducing crosslink agents with the more rapid and precise laser tool. However, the occurrence of laser-induced crosslink depends on a combination of factors, including laser source parameters, photo-reactivity of the nucleotides and type of protein. For this reason, we focus here on a rapid method to screen, verify and optimize the yield of crosslink induction, using as model breast cancer cells and a fluorescent reporter. The model is easy of use, very rapid and ensures the transferability of results and represents a good candidate to orient a pre-screening phase high throughput experiments based on L-crosslinking, as shown in the scheme in Figure 1.

Human Caucasian, breast adenocarcinoma (triple negative breast cancer-MDA-MB-231) cells were transfected with a vector coding for Estrogen Receptor $\alpha$ (ER$\alpha$) linked to Green Florescent protein (GFP). The clonal population, MDA-ER$\alpha$-GFP, stably expressed by the transcriptional factor ER$\alpha$-GFP, was selected and isolated with cell-sorting. Sorting was carried out using as control parental MDA-MB-231 cells and setting the highest threshold of GFP intensity.

We carried on tests to check whether cell transfection with the GFP fluorescent reporter was successful. In Figure 2 the expression of ER$\alpha$-GFP (A) and the fluorescence intensity of cell population are reported (B).

ER$\alpha$ is a ligand-activated nuclear receptor with binding motifs on DNA, defined as Estrogen Responsive Elements (ERE). ER$\alpha$-GFP transfected protein maintains its specificity for ERE sequences; after the crosslink induction in living cells with laser light, followed by a gentle extraction and chromatin purification, the intensity of GFP fluorescence provides the measure of the quantity (and efficacy) of stable bonds formation between DNA and ERα-GFP.

To validate the screening method, two different sets of parameters of our fs laser source were used: high repetition rate with low energy/pulse-20 kHz, E = 8μJ- and low repetition rate with high energy/pulse-200 Hz, E= 125 μJ. The target was constituted by 1 mL of physiological saline solution with $10^6$ cells. In Figure 3 the crosslinking yield is reported, as easily and quickly measured by fluorimetry, in several conditions for L-crosslinking and ordinary formaldehyde-based crosslinking. Comparing the fluorescence intensity in untreated cells (ctr-), chemically-treated for ordinary crosslinking (ctr+, formaldehyde for 20 minutes) and L-crosslinked conditions, the same crosslink yield has been obtained with few high-energy or with many low-energy pulses. This is an indirect indication that the total energy released during L-crosslink may play a role in the formation of stable bond between ERE and ERα-GFP (figure 3A) and that, in our irradiation condition, the L-crosslinking process is based on a linear excitation scheme of DNA. In our case the total delivered energy was 9.6 and 48 J for 8 μJ, 20 kHz pulses with 60 and 300 seconds of irradiation time, respectively, and 1.5 and 7.5 J for 125 μJ, 200 Hz pulses, respectively. With both combinations of repetition rate (200 Hz and 20 kHz) and energy/pulse (125 and 8 μJ), the fluorescence yield is even higher than the one obtained with the chemically-induced process, in support of L-crosslinking once again. The method is easy and fast and likely one minute of irradiation time is enough to get the maximum L-crosslink yield, as suggested by the absence of fluorescence signal by increasing about five times the irradiation time (Figure 3B). This is most likely due to saturation of the crosslinking in the cell target already in one minute of irradiation. The fast-screening fluorescence-based platform was also used to test the optimal cellular concentration to induce and detect the L-crosslink (Figure 3C). For 8 µJ, 20 kHz and 125 µJ, 200 Hz irradiation conditions, $10^6$ cells/mL was the highest and best concentration, suggesting to use an overall number of cells in the $10^6$ range.

Results reported in Figure 3 suggest that the two laser conditions i.e. 20 kHz with E=8µJ and 200 Hz with E=125µJ give nearly the same result, in terms of crosslinking yield with a delivered total dose of energy ranging between 1.5 and $\approx$ 50 µJ. Moreover, the ratio of the delivered total energy for the two laser source parameter settings, 9.6/1.5 = 6.4 would suggest that higher energy pulses are more than six times more effective than lower energy pulses in inducing L-crosslinking. However, it must be considered that in the interaction between UV light and living cells, the biological damage parameters take a key role. Therefore, we checked a number of indicators typically referred to as cell damage indicators. Amongst these indicators we also considered the chance that irradiated cells, though not immediately killed by laser irradiation, might start apoptosis after irradiation and, because of that, might be seriously altered by L-crosslinking. This was checked by waiting a 30 minutes time slot from irradiation to the protein extraction protocol. In Figure 4 a panel of protein induced by DNA damage is shown with and without the 30 minutes time slot from the L-crosslink induction to the starting of protein extraction protocol (4A), as well as the percentage of cell death following laser irradiation (4B). In fact, during the 30 minutes waiting time, biological pathways leading to cell death are activated, as proven by the increase of phS139 γH2AX and phS15 p53. These data highlight the need of a balanced combination between yield of L-crosslink and cell integrity parameters.

From the above tests, reported in Figure 3 and 4, it seems that 125 µJ, 200 Hz pulses are to be preferred to 8 µJ, 20 kHz pulses in view of L-crosslinking while minimizing the induced undesired damage. However, experiments regarding the DNA amplificability after laser irradiation in in-*vitro* tests (Russmann et al. 1997; Russmann et al. 1998) and more recent experiments regarding DNA amplificability and repair in living cells after laser treatment (Nebbioso et al. 2017), which is an essential step for ChIP analysis and high throughput screening of the fixed DNA-protein interactions in living cells, indicated that 8 µJ pulses are preferable over 125 µJ from this point of view. This is yet an indication that L-crosslinking methods need a careful evaluation of the side-way induced damages and that a careful trade-off between efficacy of the induced crosslinking and overall biological impact on the treated cells have to be considered.

*Materials and Methods*

*Cell line*

MDA-ERα-GFP were cultured in Dulbecco's modified Eagle's medium (DMEM-Euroclone) supplemented with 10% FCS (Sigma), 50mg/mL penicillin-streptomycin (Gibco) and 2mM glutamine (Gibco). Cells were maintained at 37°C in a humidified atmosphere of 95% air, 5% $CO_2$ with the adding of G418 (Invitrogen) at the final concentration of 0,5 mg/mL for transfected cells.

*Transfection of pERalpha-EGFP-C1 in MDA-MB231 cells*

pERalpha-EGFP-C1 vector (kindly provided by Ken-Ichi Matsuda, Department of Anatomy and Neurobiology, Kyoto Prefectural University of Medicine) was transfected, following the manufacturer's procedure, into MDA-MB231 cells by the use of Lipofectamine™ 2000 Transfection Reagent (Invitrogen).

*Transfection efficiency evaluation*

The efficiency of pERα-EGFP-C1 vector transfection in MDA-MB231 cells was calculated with FACS (FACScalibur; BD Biosciences, San Jose, CA). The percentage of GFP-positive cells was analyzed as shift along FL1 channel, in comparison with empty vector transfected cells.

*Cell sorting*

MDA-ERα-GFP cells were sorted with FACS ARIAII (Becton Dickinson). Data were analyzed with Diva 6.1 software. After the construction of G1 gate on the base of physical parameters (FSC and SSC) to isolate the homogeneous cell population, the dot-plot of GFP-positive cells was obtained. The FACS calibration was carried out through Accudrop beads (BD), needles and connections were sterilized with 70% ethanol and washed with sterile PBS. The threshold of GFP positive cells was set on G2 gate and the cells were collected.

*Immunoprecipitation assay (IP)*

Cell extract was prepared from MDA-ERα-GFP cells, grown at 80% confluence in 150-mm tissue culture plates. Immunoprecipitation procedure was according to previous reported (Nebbioso et. Al 2009). ERα, GFP and IgG rabbit control antibodies were purchased from Santa Cruz.

*Fluorescence Experiments*

Irradiated and control cells were gently lysed (Buffer: 20 mM HEPES pH 7.6, 10 mM EDTA, 0.5 mM EGTA, 0.25% triton-x100) for 10 minutes a 4°C to preserve the integrity of nuclear membranes. Then nuclei were sonicated and the chromatin was recovered. DNA crosslinked with ERα-GFP was purified by the binding with diethylaminoethylcellulose disc filters (DE81); DE81 filters were than extensively washed (Buffer: 50 mM HEPES pH 7.6, 150mM NaCl, 1mM EDTA, 0.5mM EGTA) for five times. GFP signal was measured (excitation = 485 nm; emission = 520nm) with fluorescence scan plate reading (TECAN INFINITE 200). Each experiment was carried out in triplicates.

*Cell death analysis*

After L-crosslink induction cells were collected in 0.1% sodium citrate and 50μg/mL Propidium Iodide and incubated for 30 minutes in the dark. The percentage of cells with sub-$G_1$ DNA was analyzed with FACS (FACScalibur; BD Biosciences, San Jose, CA).

*Protein extraction Western blotting and antibodies*

Total and histone protein extraction from MDA-ERα-GFP cells was performed as reported in (Nebbioso et al.2017). Samples concentration was determined by Bio-Rad assay (Biorad) and 30μg of total proteins and 5μg of histone extraction were loaded into SDS-PAGE minigels, transferred to nitrocellulose membranes (Schleicher and Schuell, Dassel, Germany) and recognized by phS139 H2AX (abcam) and PhS15 p53 (Upstate). Total ERKs and H1 (Santa Cruz) were used as loading controls. Primary antibodies were detected with horseradish peroxidase-linked anti-mouse or anti-rabbit conjugates (Amersham Biosciences), and were visualized using the ECL detection system (Amersham Biosciences).

*Fs Laser Source and laser parameters settings*

Our laser source and its settings for the performed experiments has been already described elsewhere (Nebbioso et al. 2017). Briefly, the device is a PHAROS-based laser system (Light Conversion), optimized to perform best in the UV domain ($\lambda < 300$ nm) at a pulse repetition rate of 2 kHz. The experiments were performed using two UV wavelengths, 258 and 300 nm, delivered by a HIRO optional wavelength converter and an ORPHEUS optical parametric amplifier, respectively. The carrier wavelength was set at 258 nm to fit within the spectral range of 250–280 nm, in which DNA base absorption is peaked. Irradiation at 258 nm causes the electronic excitation of purines and pyrimidines, triggering the formation of new bonds between neighboring molecules. Specifically, UV light generates covalent binding between the reactive bases of DNA (thymine and cytosine) and amino acids (mainly phenylalanine, cysteine and tyrosine).

**Chern-Simons theory applied to preliminary KRAS gene analysis**

As it is well known, General Relativity is the best accepted theory describing gravitational interaction, which completely changed the view of physics, giving life to several new research areas. However, it is in contrast with observations both at low and at high energies; for instance, it is still impossible to get a complete and coherent theory of Quantum Gravity, since General Relativity seems to be unable to describe the gravitational interaction at the quantum level. With regards to cosmological scales, it cannot predict the today observed universe exponential expansion as well as the early-time universe inflation, namely the expansion of the universe in the first $10^{-33}$ seconds after the Big Bang (Beringer et al. 2012). Though the best accepted explanations to these problems involve Dark Matter and Dark Energy, these dark side tools are supposed to represent the 96% of the whole universe without having been observed so far. Other possible resolutions to General Relativity issues are the alternative theories of gravity, which extend (or modify) the Hilbert-Einstein action, providing theories which well works at any scales (Capozziello and De Laurentis 2011); one of them is the

Chern-Simons gravity (Jackiw and Pi 2003).

Recently, Chern-Simons theories have been imported from the above scientific scenario to interpret complicated biological structures and configurations.

To briefly remind the application of Chern-Simons formalism in three dimensions to genetic sequence analysis (Capozziello et al. 2018; Capozziello and Pincak 2018) we recall the basic concepts of the theory, with the final aim of the schematization of the interactions between different parts of DNA or RNA.

The considered configuration space contains all the possible combinations among nitrogen bases. As nitrogen bases combine with each other in triplets, such space will be three-dimensional. We shortly outline the theoretical structure which lies behind the application to genomic sequences; a particular kind of Chern-Simons action, $S$, in three dimensions can be written as the following trace (Birmingham et al 1991):

$$S = \int \text{Tr}\left[\mathbf{A} \wedge d\mathbf{A} + \frac{2}{3}\mathbf{A} \wedge \mathbf{A} \wedge \mathbf{A}\right] \qquad (1)$$

being "$\wedge$" the external product, "d" the external derivative and $\mathbf{A}$ the one-form gauge connection. For instance, in the electromagnetic theory, under the formalism of gauge interaction, the one-form connection turns out to be the vector potential defined as $\nabla \times \mathbf{A} = \mathbf{B}$. The only gauge invariant measurable quantity coming from the above action is the so-called Wilson Loop (Maldacena 1998):

$$W(\mathbf{A}) = \text{Tr}[\text{Exp}\{i\text{P} \oint \mathbf{A}\}] \qquad (2)$$

where P stands for the factor ordering; the expectation value of the Wilson Loop provides the *Chern-Simons current "j"*.

With regards to our applications, in order to get the connection for the Chern-Simons action in the considered space of configuration, we may properly define our nitrogen bases as a set of four quaternions and, subsequently, perform all the necessary calculations to get the Chern-Simons current for all possible triplets. We define the four Nitrogen bases on the DNA as a set of quaternions depending on the gene $\beta_i$:

$$A_{DNA} = \left[e^{i\frac{\pi}{2}\beta_i}\right] + [0]\boldsymbol{i} + [0]\boldsymbol{j} + [0]\boldsymbol{k}$$

$$T_{DNA} = [0] + \left[e^{-i\frac{\pi}{2}\beta_i}\right]\boldsymbol{i} + [0]\boldsymbol{j} + [0]\boldsymbol{k} \qquad (3)$$

$$C_{DNA} = [0] + [0]\boldsymbol{i} + \left[e^{i\pi\beta_i}\right]\boldsymbol{j} + [0]\boldsymbol{k}$$

$$G_{DNA} = 0 + [0]\boldsymbol{i} + [0]\boldsymbol{j} + \left[e^{i\,2\pi\beta_i}\right]\boldsymbol{k}$$

and the same holds for RNA. Through the above definition of DNA in terms of quaternions, we also define a set of 8 states $s_l$ ($l$ runs from 1 to 8) as follows:

$$[s_1] = ([A],[T^*]) \quad [s_2] = ([A],[T]) \quad [s_3] = ([C],[G^*]) \quad [s_4] = ([C],[G])$$

$$[s_5] = ([T],[T^*]) \quad [s_6] = ([T],[T]) \quad [s_7] = ([G],[G^*]) \quad [s_8] = ([G],[G])$$

If one wants to study the docking DNA-RNA, we can take as a connection the Christoffel symbol $(A_i)_k^j$, so that after defining the metric tensor as the average of the tensorial product $g_{ij} = <[s_i],[s_j]^*>$, the metricity principle $\nabla g_{ij} = 0$ is immediately recovered.

However, in the computation of Chern-Simons current for the genetic code, the one-form connection can be simply labeled by the nitrogen basis itself so that, considering eq. (2) and defining the Chern-Simons current as the expectation value of the Wilson Loop $j = <W(A)>$, we get the Chern-Simons currents for all the possible triplets of the genetic code. In this way, the 64 components of the metric tensor turns into the 64 triplets combinations of the genetic code.

Without going through further details that are reported elsewhere in (Capozziello et al. 2018; Capozziello and Pincak 2018), we report the list of the Chern-Simons current value associated to each of the possible 64 triplets of basis of the genetic code in the Table 1. As in the space of coordinates the Chern-Simons currents are linked to the spacetime curvature, in the space of nitrogen bases they represent a point-like curvature of a certain string, to be compared with the curvature of a close-by amino acid or with another mutated sequence in order to check, for example, whether docking between two biomolecules is likely or not. As a matter of example, the docking between V3 loop region of HIV with the *CD4* gene of the host cell has been investigated, by deriving a master equation to study the attach of the two molecules (Capozziello et al. 2018). In fact, the computation of Chern-Simons currents may give information about the docking point or, at least, about the point with the highest probability of anchoring, where we expect to find the same current for both sequences. These points or regions will be the domains of the two molecules that best match the curvature from both sides.

Here we decided to apply the Chern-Simons theory to study the mutations by a single base replacement of a selected region in the KRAS human gene, to possibly state whether the studied mutations are likely or not.

In particular, we focus on the string "chr12: 25,380,173-25,380,346" whose triplets are shown as the top rows in Table 2 together with its mutated sequence taken by the BIOMUTA database, displayed in the corresponding bottom rows of Table 2. In this case all the variations are single-nucleotide, within the considered triplet, as reported in the database (https://hive.biochemistry.gwu.edu/biomuta) where human pathologies associated to each single mutation can also be checked. For the sake of clarity we summarize the only variations of the sequence in Table 3.

We calculated the Chern-Simons current value associated to each of the two sequences according to Table 1, whose behavior is shown in Figure 5.

We interpreted the results based on the following criterion: mutated sequence, containing point-like variations far from the original sequence, step away from the equilibrium state and, therefore, can be considered less likely than those not causing significant changes. A quantitative measurement of the significance of a specific mutation is represented by its change in the Chern-Simons current value, as compared to that calculated for the original sequence. That being said, an abrupt variation of the current is supposed to indicate an unlikely mutation, while those mutations which have a larger probability to occur, are characterized by a smooth variation. This is pointed out in depth in (Bajardi et al. 2020) where the authors focus on different parts of the same gene, comparing original and mutated sequences in order to investigate the biological implications of this new method.

In Figure 5 the black-solid and red-dashed lines refer to the original and the mutated sequences, respectively. Eight mutations are present in the mutated sequence as specified in Table 3. The most significant mutations are located at the position 22, 31 and 34 (blue highlight in Figure 5), which are characterized by the highest change in the corresponding Chern-Simons values. Because of that, they significantly change the trend of the graph and can be interpreted as unlikely or infrequent. It is worth checking on the BIOMUTA database what disease the above three mutations are associated to. In particular, the mutation at position 22 implies brain cancer and malignant glioma, the mutation at position 31 implies thyroid carcinoma, and the mutation at position 34 leads to uterine cancer.

It is worth stressing out that the approach outlined in this section represents a new method still under investigation. Here we just wanted to give a general view on the formalism and on its applications, without claim of completeness, presenting its main features in the application to analyze base sequences in nucleic acids. In a future prospect, it might be also possible to study the docking between different biomolecules. By knowing the DNA/RNA sequences of two interacting molecules, one can work out the corresponding Chern-Simons current and check out the difference of curvature triplet by triplet. We expect that, in analogy to what happens in the real spacetime, those parts with similar curvature will be expected to be those where the docking will occur. On the contrary, where the Chern-Simons currents will yield very different curvature, the docking will not take place. As an example, one might apply the above method to analyze the docking interaction between ERα and GFP.

**Conclusions**

In conclusion we presented two different methods, based on very recent concepts and technologies deeply intrinsic to physics, that apply to up-to-date hot topics of molecular biophysics and genetics, such as (i) techniques to fix interactions between nucleic acids and proteins in living cells with no alteration of the bio-chemical equilibrium and (ii) methods to quantitatively analyze bio-molecules configuration, potentially capable to describe their mutual complex interaction, such as docking between proteins and nucleic acids. These methods, very much far from each other, are proposed within a multi-disciplinary scenario to show how concepts and methods of physics, even very much far apart, can enter bio-sciences to yield a significant progress in understanding the inherent complexity of hot topics such as the ones above mentioned.

One of the main problems in probing interactions between biomolecules, such as DNA-protein interplay, is caused by the impossibility of chemical crosslinkers to select direct versus indirect bindings or short-lived chromatin occupancy. A recent technique, characterized by crosslinking cells by UV laser pulse irradiation, has been considered. This technique, combined to ChIP in high throughput experiments, calls for a pre-screening fast method in order to set up suitable irradiation conditions of the cell target for effective L-crosslinking without final and lengthy ChIP analysis.

A fast pre-screening method has been here presented in detail, where living human cells have been first transfected with a vector linked to Green Florescent protein (ERα-GFP), so that the well-known interaction between the Estrogen Receptor Elements (ERE) region of the cell DNA and the ERα protein can be simply tagged by studying the fluorometric response of the irradiated cells. The biological impact to cells by UV irradiation is investigated by looking at DNA integrity, proteins stability and cellular viability.

A latter novel and different approach is presented to analyze or re-visit DNA and RNA sequences; this approach, in principle, might be used with the aim to study the behavior of biomolecules to determine their configuration and the role played by this configuration in reciprocal interactions such as docking. It is based on methods derived from Chern-Simons super-gravity, suitably adapted to describe interactions between nucleic acids. As a preliminary test case, in order to introduce the method features, we analyzed a fairly small region of KRAS human gene sequence and some of its mutations by substitution of single bases. Our model is capable to identify and possibly predict the position of mutations within a sequence by analyzing the values of the Chern-Simons current, defined as the expectation value of the so-called Wilson loop, the only gauge invariant measurable quantity of the theory. The prospect of predicting where possible mutations may be located in a given genetic sequence, together with their likelihood, definitely represent an exciting potential progress and powerful tool for sequences analysis, that will be strongly investigated in future studies.

C.A. and F.B. aknowledge the project PON (Programma Operativo Nazionale Ricerca e Innovazione) 2014-2020 (CCI 2014IT16M2OP005), "Tecnologie innovative per lo studio di interazioni tra acidi nucleici e proteine: metodi sperimentali e modelli"; project code DOT1318991. C.A., M.V. and M.S. acknowledge support from the Italian Ministry for Research under the Project PRIN - Predicting and controlling the fate of bio-molecules driven by extreme-ultraviolet radiation - Prot. Nr.20173B72NB. Most of the authors acknowledge: EU project ATLAS (221952). L.A. acknowledges MIUR20152TE5PK; EPICHEMBIO CM1406; AIRC-17217; VALERE: Vanvitelli per la Ricerca; Campania Regional Government Technology Platform Lotta alle Patologie Oncologiche: iCURE; Campania Regional Government FASE2: IDEAL. MIUR, proof of concept, CUP:B64I19000290008

Table 1. Value for the Chern-Simons current, which is an adimensional positive number between 0 and 1, associated to each possible of the $4^3$ base triplets (Capozziello et al. 2018), with Adenine (A), Thymine (T), Guanine (G), Cytosine (C). Reproduced with permission from Capozziello 2017 by Annalen Der Physik.

| Amino acid | Chern-Simons Current | Amino acid | Chern-Simons Current |
|---|---|---|---|
| Phe (TTT) | j = 0.7071 | Tyr (TAT) | j = 0.0214 |
| Phe (TTC) | j = 0.5000 | Tyr (TAC) | j = 0.0205 |
| Leu (TTA) | j = 0.3717 | Sto (TAA) | j = 0.0197 |
| Leu (TTG) | j = 0.2887 | Sto (TAG) | j = 0.0189 |
| Leu (CTT) | j = 0.2319 | His (CAT) | j = 0.0182 |
| Leu (CTC) | j = 0.1913 | His (CAC) | j = 0.0175 |
| Leu (CTA) | j = 0.1612 | Gin (CAA) | j = 0.0169 |
| Leu (CTG) | j = 0.1382 | Gin (CAG) | j = 0.0163 |
| Ile (ATT) | j = 0.1201 | Asn (AAT) | j = 0.0157 |
| Ile (ATC) | j = 0.1057 | Asn (AAC) | j = 0.0152 |
| Ile (ATA) | j = 0.0939 | Lys (AAA) | j = 0.0147 |
| Met (ATG) | j = 0.0841 | Lys (AAG) | j = 0.0142 |
| Val (GTT) | j = 0.0759 | Asp (GAT) | j = 0.0138 |
| Val (GTC) | j = 0.0690 | Asp (GAC) | j = 0.0134 |
| Val (GTA) | j = 0.0630 | Glu (GAA) | j = 0.0129 |
| Val (GTG) | j = 0.0579 | Glu (GAG) | j = 0.0126 |
| Ser (TCT) | j = 0.0534 | Cys (TGT) | j = 0.0122 |
| Ser (TCC) | j = 0.0495 | Cys (TGC) | j = 0.0118 |
| Ser (TCA) | j = 0.0460 | Sto (TGA) | j = 0.0115 |
| Ser (TCG) | j = 0.0429 | Trp (TGG) | j = 0.0112 |
| Pro (CCT) | j = 0.0402 | Arg (CGT) | j = 0.0109 |
| Pro (CCC) | j = 0.0377 | Arg (CGC) | j = 0.0106 |
| Pro (CCA) | j = 0.0354 | Arg (CGA) | j = 0.0103 |
| Pro (CCG) | j = 0.0334 | Arg (CGG) | j = 0.0100 |
| Thr (ACT) | j = 0.0316 | Ser (AGT) | j = 0.0098 |
| Thr (ACC) | j = 0.0299 | Ser (AGC) | j = 0.0096 |
| Thr (ACA) | j = 0.0284 | Arg (AGA) | j = 0.0093 |
| Thr (ACG) | j = 0.0270 | Arg (AGG) | j = 0.0091 |
| Ala (GCT) | j = 0.0257 | Gly (GGT) | j = 0.0089 |
| Ala (GCC) | j = 0.0245 | Gly (GGC) | j = 0.0087 |
| Ala (GCA) | j = 0.0234 | Gly (GGA) | j = 0.0085 |
| Ala (GCG) | j = 0.0224 | Gly (GGG) | j = 0.0083 |

Table 2. Sequence of the string "chr12: 25,380,173-25,380,346" in the KRAS human gene with its mutated version as taken from the BIOMUTA database (https://hive.biochemistry.gwu.edu/biomuta). The original sequence, a word of 173 bases, is reported in the top rows with the mutated one in the bottom, in a one-to-one correspondence base-by-base. Triplets, that code for amino acids, are separated by thicker lines. Mutated bases are indicated in red in the mutated sequence.

| A | T | G | G | T | G | A | A | T | A | T | C | T | T | C | A | A | A | T | G | A | T | T | T | A | G | T | A | T | T | A | T | T | T | A | T | G | G | C | A | A | A | T | A | C |
|---|---|---|---|---|---|---|---|---|---|---|---|---|---|---|---|---|---|---|---|---|---|---|---|---|---|---|---|---|---|---|---|---|---|---|---|---|---|---|---|---|---|---|---|---|
| A | T | G | G | T | G | A | A | T | A | T | **T** | T | T | C | A | A | A | T | G | A | T | T | T | A | G | T | A | T | T | A | T | T | **C** | A | T | G | G | C | A | A | A | T | A | C |

| A | C | A | A | A | G | A | A | A | G | C | C | C | T | C | C | C | C | A | G | T | C | C | T | C | A | T | G | T | A | C | T | G | G | T | C | C | C | T | C | A | T |
|---|---|---|---|---|---|---|---|---|---|---|---|---|---|---|---|---|---|---|---|---|---|---|---|---|---|---|---|---|---|---|---|---|---|---|---|---|---|---|---|---|---|
| A | C | A | A | A | G | A | A | A | G | C | C | C | T | C | C | C | C | A | G | T | C | **A** | T | C | A | T | G | T | A | C | T | G | G | T | C | C | C | T | C | A | T |

| T | G | C | A | C | T | G | T | A | C | T | C | C | T | C | T | T | G | A | C | C | T | G | C | T | G | T | G | T | C | G | A | G | A | A | T | A | T | C | C | A | A | G | A | G |
|---|---|---|---|---|---|---|---|---|---|---|---|---|---|---|---|---|---|---|---|---|---|---|---|---|---|---|---|---|---|---|---|---|---|---|---|---|---|---|---|---|---|---|---|---|
| T | G | C | A | **T** | T | G | T | A | C | T | C | **A** | T | C | T | T | G | A | C | C | T | **A** | C | T | G | T | G | T | C | G | A | G | A | A | T | A | T | C | C | A | A | G | A | G |

| A | C | A | G | G | T | T | T | C | T | C | C | A | T | C | A | A | T | T | A | C | T | A | C | T | T | G | C | T | T | C | C | T | G | T | A | G | G | A | A | T | C |
|---|---|---|---|---|---|---|---|---|---|---|---|---|---|---|---|---|---|---|---|---|---|---|---|---|---|---|---|---|---|---|---|---|---|---|---|---|---|---|---|---|---|
| A | C | **G** | G | G | T | T | T | C | T | C | C | A | T | C | A | A | T | T | A | C | T | A | C | T | T | G | C | T | T | C | C | T | G | T | A | G | G | A | A | T | **T** |

Table 3. Mutations in the "chr12: 25,380,173-25,380,346" KRAS gene as reported by the BIOMUTA database (https://hive.biochemistry.gwu.edu/biomuta).

25,380,346 C -> T
25,380,307 A -> G
25,380,282 G -> A
25,380,272 C -> A
25,380,264 C -> T
25,380,240 C -> A
25,380,206 T -> C
25,380,184 C -> A

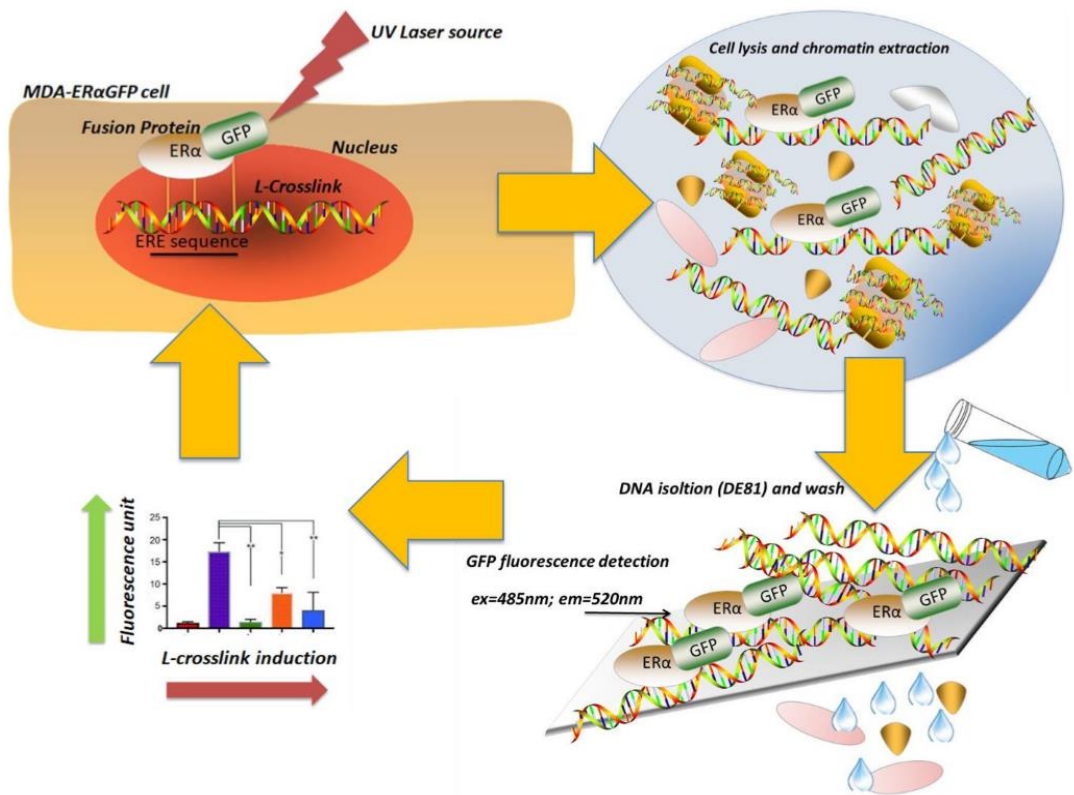

Figure 1. Scheme of the pre-screening method.

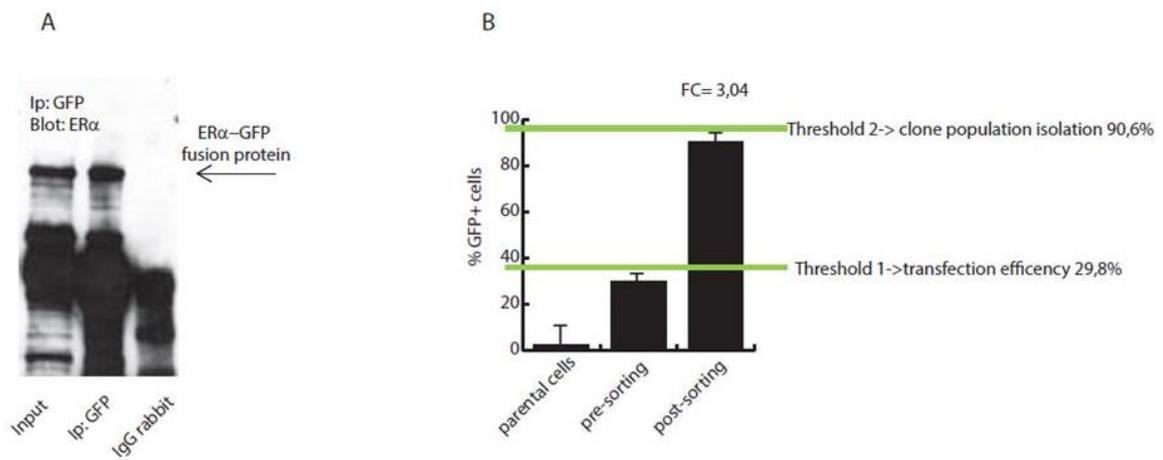

Figure 2. Diagnostics of the transfected cell line. Expression of ERα-GFP (Nebbioso et al. 2017) (A) and fluorescence intensity of cell population (B) with successful final cell sorting of the fluorescent-engineered cells with a clone isolation population higher than 90%.

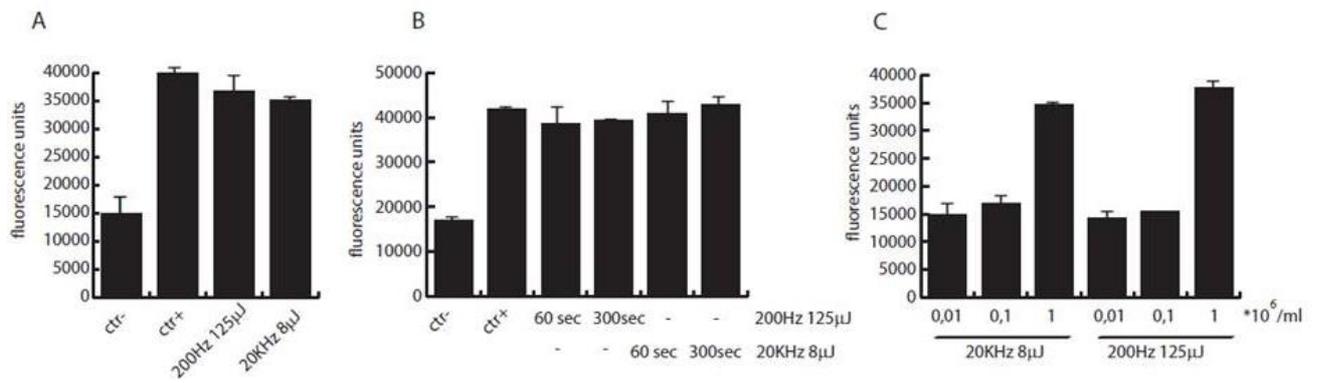

Figure 3. Study of the crosslinking yield. (A) Comparison of the method efficacy for L-crosslinking with (200 Hz, 125 µJ), (20 kHz, 8 µJ) pulses, irradiation time of 60 seconds, and ordinary chemical crosslinking. (B) Comparison of the method efficacy for different irradiation times in L-crosslinking. Ctr+ and Ctr- stand for treated (chemically-treated formaldehyde for 20 minutes) and untreated cells, respectively. (C) Comparison of the method efficacy for different values of cell concentration.

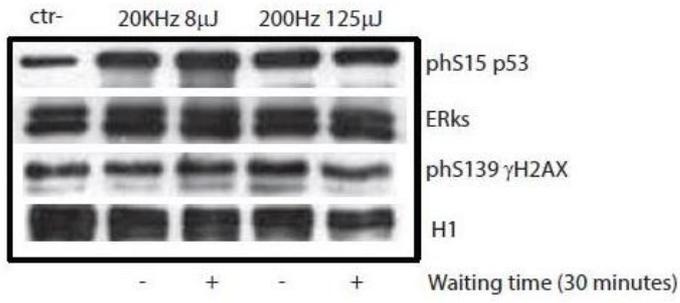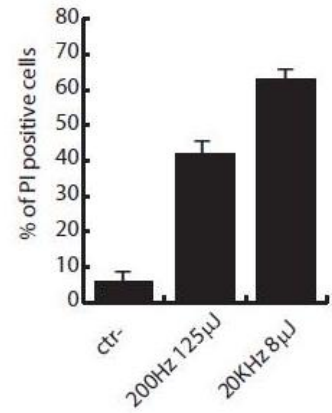

Figure 4. Evaluation of cell and DNA damage due to L-crosslinking. (A) Activation of cell death pathways revealed by the increase of the phS139 γH2AX and phS15 p53 proteins. (B) Direct evaluation of the cell death percentage following laser irradiation. In both figures, Ctr- stands for untreated cells.

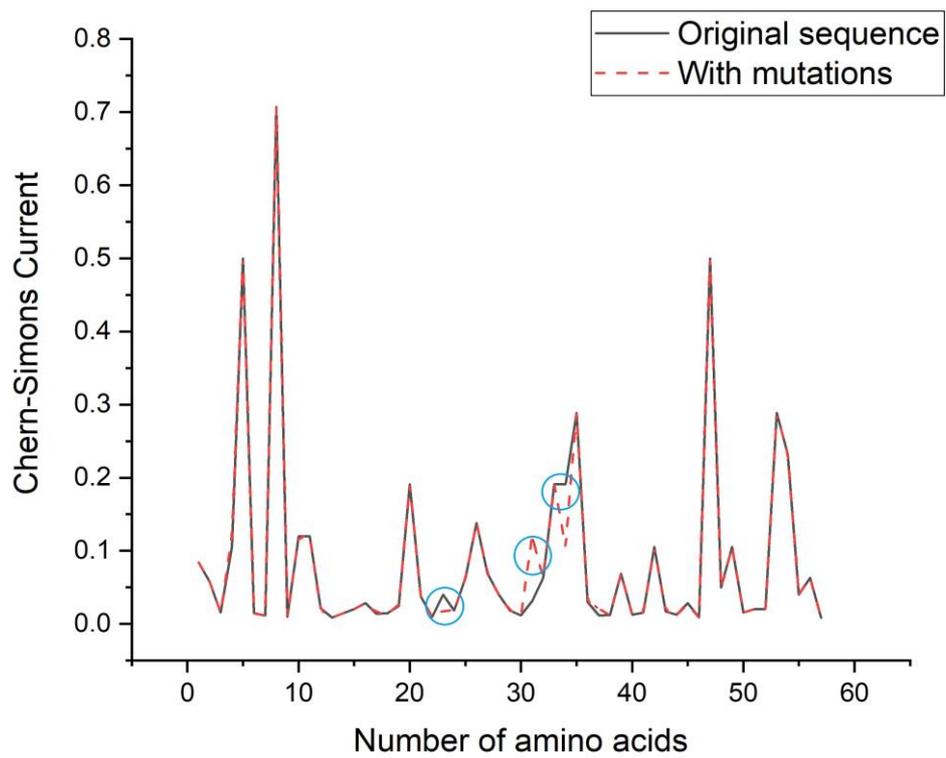

Figure 5. Chern-Simons current value for the original chr12: 25,380,173-25,380,346" KRAS sequence (black-solid) and its mutated version (red-dashed) versus the number of amino acids, each one corresponding to a base triplet. The main point-like mutations are highlighted by blue circles and located in the amino acid position 22, 31 and 34.